\RequirePackage{fix-cm}
\documentclass[smallextended]{svjour3}       
\smartqed  
\usepackage{graphicx}
\usepackage{amsmath,amssymb}
\usepackage{url}
\begin{document}

\title{Constraints on Kaluza-Klein Gravity\\ from Gravity Probe B
}


\author{J.M. Overduin \and R.D. Everett \and P.S. Wesson 
}


\institute{J.M. Overduin and R.D. Everett \at
              Department of Physics, Astronomy and Geosciences \\
              Towson University, Towson, MD, 21252, U.S.A. \\
              Tel.: +1-410-704-3624\\
              \email{joverduin@towson.edu}           
           \and
           P.S. Wesson \at
              Department of Physics and Astronomy \\
              University of Waterloo, Waterloo, ON, N2L 3G1, Canada \\
              \email{psw.papers@yahoo.ca}
}

\date{Received: date / Accepted: date}

\maketitle

\begin{abstract}
Using measurements of geodetic precession from Gravity Probe B, we constrain possible departures from Einstein's General Relativity for a spinning test body in Kaluza-Klein gravity with one additional space dimension. We consider the two known static and spherically symmetric solutions of the 5D field equations (the soliton and canonical metrics) and obtain new limits on the free parameters associated with each. The theory is consistent with observation but must be ``close to 4D'' in both cases.

\end{abstract}

\section{Introduction}

There is now a substantial literature on the higher-dimensional extension
of Einstein's general theory of relativity known as Kaluza-Klein gravity
\cite{OW97,W06}.  There are several ways to test the theory,
with perhaps the most straightforward involving the motion of test objects
in the field of a static, spherically-symmetric mass like the Sun or the
Earth.  Birkhoff's theorem in the usual sense does not hold in higher
dimensions \cite{BM95,KG08,S12}, so some question arises about the best
choice of metric to describe such a situation. But the gravitational field of
the Sun must be close to Schwarzschild, by the solar system tests.  Also,
the non-linearity of Einstein's equations requires us to use exact solutions
to model the field. In the minimal five-dimensional (5D) case, only two such
are known: the soliton \cite{S83,GP83,DO85} and canonical solutions
\cite{MLW94,LW96a,MWL98}.  Both satisfy the vacuum field equations in 5D,
consistent with the spirit of Kaluza's original idea that 4D matter and
gauge fields appear as a manifestation of pure geometry in the
higher-dimensional world. The soliton metric contains no explicit
$\ell$-dependence and reduces to the standard 4D Schwarzschild solution
on hypersurfaces $\ell=$~const.  The canonical metric contains explicit
$\ell$-dependence; but its effects are suppressed by a quadratic prefactor
$(\ell/L)^2$ where $L$ is a constant and presumably a large length scale
(a free parameter of the theory). The fifth dimension in this solution 
is flat.

The classical tests of general relativity were first applied to the soliton
by Kalligas et al. and Lim et al. in 1995 \cite{KWE95,LOW95}, and again in
more generality by Liu and Overduin in 2000 \cite{LO00}. Light deflection,
perihelion precession of Mercury, and radar ranging to Mars set limits of
order $10^{-2}$ on the primary free parameter of the metric, and the hope
was expressed that data on geodetic precession for spinning test masses 
from Gravity Probe~B might push this down to as little as $10^{-4}$.
Overduin then used observational constraints on violations of the
equivalence principle to obtain bounds of order $10^{-6}-10^{-8}$ on the
soliton metric as applied to the Sun, Earth, Moon and Jupiter \cite{O00}.
If the soliton metric were the only choice available,
such a result might call the need
for a higher-dimensional extension of general relativity into question,
contrary to what is suggested by most attempts to unify gravitation with
the other fundamental interactions. However, analysis of the canonical
metric by Mashhoon, Wesson and Liu \cite{MWL98} has revealed that for this
solution, the classical tests are satisfied {\em exactly\/} for non-spinning
test bodies. This is due to the flatness of the extra dimension and the
quadratic prefactor on the 4D part of the metric and can be proven using
a 1926 theorem on embedding by Campbell; see Ref.~\cite{W11a} for discussion.

Spin thus emerges as a potentially critical discriminator
between standard and higher-dimensional extensions of general relativity,
and the canonical solution may be the most appropriate for this problem.
The geodetic effect for the canonical metric was first worked out by Liu and
Wesson in 1996 \cite{LW96a}. We return to this work and assess the status of
the theory using the recently released final results from Gravity
Probe~B (henceforth GPB \cite{E11}). The geodetic effect is briefly
reviewed in Section~2. We consider the soliton metric in Section~3
and the canonical metric in Section~4. Our results are summarized
and discussed in Section~5.

\section{Geodetic effect}

\begin{figure*}
\includegraphics[width=1.0\textwidth]{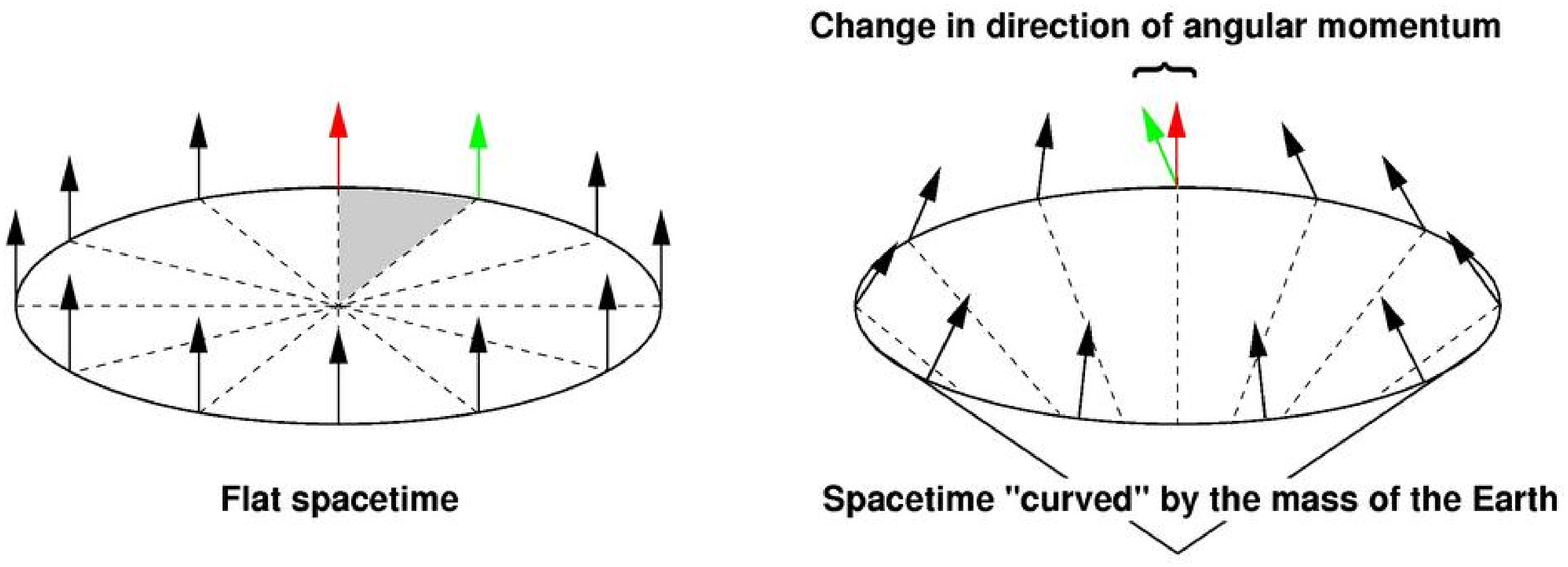}
\caption{The ``missing inch'' model for geodetic precession. A gyroscope's spin vector (arrow) is orthogonal to the plane of its motion, and in flat spacetime its direction is unchanged as the gyroscope completes an orbit. If, however, space is folded into a cone to simulate curvature due to the earth (right), then part of the area inside the circle (shaded at left) must be removed and the gyroscope's spin vector no longer lines up with itself after making a complete circuit. This angular shift contributes two-thirds of the total geodetic effect, and the difference between the circumference of the orbit with and without this effect at GPB's operating altitude of 642~km is about an inch \cite{T88}.}
\label{fig:missingInch}       
\end{figure*}

The geodetic effect is the first test of general relativity to involve the
spin of the test body (the other being the frame-dragging or Lense-Thirring
effect) and was originally investigated by Willem de~Sitter in 1916 using the
orbital angular momentum of the Earth-Moon system as a ``gyroscope'' in the
field of the Sun (this is now termed the ``solar geodetic'' or de~Sitter
effect and was also observed by GPB \cite{E88}). De~Sitter's
calculations were extended to include spin angular momentum of rotating
test bodies such as the earth by Jan Schouten in 1918
and Adriaan Fokker in 1920.
Thorne \cite{T88} has shown how the effect may be thought of as arising from
two separate contributions: one due to space curvature around the central
mass and the other a spin-orbit coupling between the spin of the gyroscope
and the ``mass current'' of the central mass (in the rest frame of the
orbiting gyroscope). The space curvature effect is twice as large as the
spin-orbit one. It arises because the gyroscope's angular momentum vector,
orthogonal to the plane of the motion, no longer lines up with itself after
one complete circuit through curved spacetime around the central mass
(Fig.~\ref{fig:missingInch}).
%
The spin-orbit contribution is analogous to Thomas precession in 
classical electromagnetism (where the electron experiences an induced magnetic
field due to the motion of the nucleus relative to its own rest frame).
Both contributions together produce what is known as the geodetic effect
within general relativity, causing the gyroscope's spin axis to precess
by a predicted 6606~milliarcseconds (mas) per year in the North-South
direction for a polar orbit at GPB's altitude of 642~km
\cite{S60,AS00,W03}.


To treat this problem in extended theories of gravity, one begins with an
appropriate choice of metric and solves the equations of motion for the
velocity (geodesic equation) and angular momentum (parallel transport
equation) of the test body, assuming that the two vectors are orthogonal.
The geodetic effect is the excess of the test body's spin angular velocity
over its orbital angular velocity.
Our notation follows that in Refs.~\cite{LW96a,LO00} except that we restore
physical units and label the extra coordinate $x^4=\ell$.
Lowercase Greek indices $\alpha,\beta,...$ run over $0,1,2,3$ as usual
while uppercase Latin indices $A,B,...$ run over all five indices $0-4$.
Proper distance in 5D ($dS$) is related to its 4D counterpart ($ds$) by
$dS^{\, 2} = ds^2 + g_{\ell\ell} \, d\ell^2$ so that
$d/dS = (ds/dS)d/ds = \sqrt{1 - g_{\ell\ell}(d\ell/dS)^2}\,d/ds$.

\section{Soliton Metric}

The line element for the soliton reads (following Ref.~\cite{GP83} but switching
to non-isotropic form and defining $a\equiv 1/\alpha$, $b\equiv \beta/\alpha$
and $M\equiv 2m$):
\begin{equation}
dS^2 = A^a c^2 dt^2 - A^{-a-b} dr^2 - A^{1-a-b} r^2 (d\theta^2 + \sin^2\theta \phi^2) - A^b d\ell^2 \; ,
\end{equation}
where $A(r)\equiv 1 - 2GM/c^2 r$, $M=M_g/a$ where $M_g$ is the gravitational
or Tolman-Whittaker mass of the soliton at $r=\infty$, and $a$ and $b$ are
free parameters related by a consistency relationship $a^2 + ab + b^2 = 1$
that follows from the field equations.
We take $b$ as the primary free parameter of the theory in what follows,
noting that the 4D Schwarzschild metric is recovered on hypersurfaces
$\ell=$~const in the limit $b \rightarrow 0$ (and $a \rightarrow +1$).
The consistency relation imposes an upper limit of
$|b| \le 2/\sqrt{3} \approx 1.15$. The 4D induced matter associated with
this solution has a density proportional to $-ab$ \cite{W11a},\footnote{The
   properties of the induced matter are obtained by decomposing the 5D field
   equations $R_{AB}=0$ into $\alpha\beta$-, $\alpha\ell$- and
   $\ell\ell$-components.  Requiring that the 4D field equations take their
   usual form, $G_{\alpha\beta}=(8\pi G/c^4)T_{\alpha\beta}$, one obtains an
   expression for the energy-momentum tensor $T_{\alpha\beta}$ of an induced 
   4D matter fluid that is a manifestation of pure geometry in 5D.}
so $a$ and $b$ must have opposite signs. Thus we are restricted
{\it a priori\/} to values of $b$ in the range $-1\lesssim b\leq 0$.

The motion of a spinning test body with
velocity $u^C\equiv dx^C/dS$ and angular momentum $S^{\, C}$ is governed by
three central equations; namely, the geodesic equation
\begin{equation}
\frac{d^2x^C}{dS^{\, 2}} + \Gamma_{AB}^C \, u^A \, u^B = 0 \; ,
\label{eq:geodesic}
\end{equation}
the parallel transport equation
\begin{equation}
\frac{dS^{\, C}}{dS} + \Gamma_{AB}^C \, S^A\, u^B = 0 \; ,
\label{eq:parallel}
\end{equation}
and the orthogonality condition
\begin{equation}
u^C \, S_C = 0 \; .
\label{eq:orthogonal}
\end{equation}
These equations can be solved analytically without placing any restrictions on
the components of $S^C$ if the orbit is taken to be circular ($\theta=\pi/2$,
$\dot{\theta}=\dot{r}=0$) \cite{LO00}.  In the weak-field limit
(i.e., dropping terms of second and higher order in $GM/c^2 r$)
the spatial part of $S^C$ is found to precess with an angular speed
\begin{equation}
\Omega=\sqrt{\frac{aGM}{c r_0^3}} \left[ 1 + \frac{3GM}{2c^2 r_0} 
\left(1-a-b\right)\right] \; ,
\end{equation}
where $r_0$ is the distance between the test body (gyroscope) and the
central mass (Earth).\footnote{For simplicity we have set to zero a constant
   of the motion ($k$ in \cite{LO00}) associated with momentum along the
   extra dimension. This has the effect of ``switching off'' the 
   $S^{\ell}$-component for the soliton metric and might be worth
   revisiting in future work.}
The geodetic effect is the excess of $\Omega$ over the test body's
{\em orbital\/} angular speed $\omega\equiv d\phi/dS$, which is found
in the same limit to be
\begin{equation}
\omega=\sqrt{\frac{aGM}{c r_0^3}} \left[ 1 + \frac{GM}{2c^2 r_0} 
\left(3-b\right)\right] \; .
\end{equation}
The accumulated geodetic precession angle per orbit,
$\delta\phi=2\pi(\omega-\Omega)/\omega$, is thus given to leading order by
\cite{LO00}
\begin{equation}
\delta\phi = \frac{3\pi GM}{c^2 r_0} \left(1 + \Delta\right) \; ,
\label{eq:precession}
\end{equation}
where the term in front of the parentheses on the left-hand side is the
standard expression for geodetic precession in 4D GR, and
\begin{equation}
\Delta = a + \frac{2}{3}b \approx \frac{b}{6} \; ,
\end{equation}
is the predicted departure in 5D theory (in the last step, we have applied
the above-noted consistency relation between $a$ and $b$). Because the free
parameter $b$ is restricted to negative values, Kaluza-Klein gravity with the
soliton metric can only accommodate precession rates {\em less than\/}
those predicted by 4D GR. This is a common feature of most attempts to
extend Einstein's theory with additional degrees of freedom, whether in the
guise of extra dimensions or new scalar fields. Physically, this is consistent
with the expectation that allowing the spin vector to wander into new regions
of dynamical phase space can only reduce (not increase) the amount of
precession that is observable in 4D.  A conclusive experimental determination
that $\Delta>0$ might therefore be the basis for ruling out the theory.
The GPB final results and implications for $\Delta$ and $b$ are
presented and discussed in Table~1 (Section~\ref{sec:discussion}).

\section{Canonical metric}

The line element for the canonical metric reads \cite{MLW94,LW96a}:
\begin{equation}
dS^{2}=\frac{\ell^{2}}{L^{2}}\biggl[B\,c^2 dt^2 - B^{-1}dr^2 - r^2
\left(d\theta^2 + \sin^2\theta d\phi^2\right)\biggr] -d\ell^2 \; ,
\label{eq:canonicalMetric}
\end{equation}
where $B(r)\equiv 1 - 2GM/c^2 r - r^2/L^2$ and $L$ is a constant length
scale, perhaps related to the cosmological constant via $\Lambda = 3/L^2$.
As before, the equations of motions 
(\ref{eq:geodesic}-\ref{eq:orthogonal}) can be solved analytically without
placing any restrictions on $S^A$ if we assume a circular orbit ($\theta=\pi/2,
u^r=u^{\theta}=0$ and model the GPB situation by placing the spin vector
in the orbital plane, $S^{\theta}=0$ with $r S^{\phi}\ll S^r$ (due to the
choice of coordinates $S^t,S^r$ and $S^{\ell}$ are dimensionless while
$S^{\theta}$ and $S^{\phi}$ have dimensions of inverse length).

Carrying out this procedure in the same way as for the soliton metric,
the spin angular velocity is found as
\begin{equation}
\Omega = \frac{c}{r_0} \sqrt{\frac{GM}{c^2 r_0} - \frac{r_0^2}{L^2}} \; .
\end{equation}
Comparing to the orbital angular velocity, and using the metric to relate
5D and 4D proper distance, one finds \cite{LW96a} that the geodetic precession
angle per orbit is again approximated in the weak-field limit
($r^2/L^2 \ll GM/c^2 r \ll 1$) by Eq.~(\ref{eq:precession}), but now with
\begin{equation}
\Delta = -\frac{2c^2r_0^2{\cal H}}{3GML} \; ,
\end{equation}
where ${\cal H}\equiv H_5/H_1\cosh[(s_0-s_m)/L]$ is a dimensionless combination 
of the normalized amplitudes of the spin vector
$[S_A S^A = -(H_t^2 + H_r^2 + H_{\ell}^2)]$ as well as the length scale $L$
and two fiducial values of the 4D proper distance. 

We do not have definitive values for all these parameters, but there are two
clear routes to testing the theory. First, we can adopt the natural assumption
that $L$ is in fact related to the cosmological constant by $\Lambda=3/L^2$
\cite{MLW94}. In observational cosmology it is usual to express $\Lambda$
in terms of the normalized dark-energy density $\Omega_{\Lambda} \equiv
\rho_{\Lambda}/\rho_{\mbox{\tiny crit}}$ 
where $\rho_{\Lambda}=\Lambda c^2/(8\pi G)$ and
$\rho_{\mbox{\tiny crit}}=3H_0^2/(8\pi G)$.
When this is done, we can use experimental constraints on $\Delta$ to put
an upper limit on the value of ${\cal H}$:
\begin{equation}
{\cal H} = -\frac{3GM}{2cH_0\sqrt{\Omega_{\Lambda}}r_0^2} \, \Delta \; .
\end{equation}
Since ${\cal H}$ is a dimensionless combination of a $\cosh$ term and a ratio
of spin components, we expect it to have a positive value not too far from
unity. 

Alternatively, we can {\em assume\/} that ${\cal H}$ is of order unity on
naturalness grounds, and use the same relationship plus experimental constraints
on $\Delta$ to put a {\em lower\/} limit on the value of the length scale
$L$:
\begin{equation}
L = -\frac{2c^2r_0^2}{3GM} \, \frac{1}{\Delta} \; .
\end{equation}
Again, we would expect that $L$ corresponds to a large distance, since the
metric~(\ref{eq:canonicalMetric}) deviates from 4D Schwarzschild by terms
of order $\ell^2/L^2$. Its sign must also be positive (since it is a 
distance).

\section{Summary and discussion}
\label{sec:discussion}

Constraints on $\Delta$ from GPB and implications for $b, {\cal H}$ and $L$ are
presented in Table~\ref{table:results}, where
\begin{table}
\caption{GPB constraints on 5D metric parameters}
\label{table:results}       
\begin{tabular}{llllll}
\hline\noalign{\smallskip}
Gyro & $r_{\mbox{\tiny NS}}$ (mas/yr) \cite{E11} & $\Delta (\times 10^{-3})$ & $|b|_{\mbox{\tiny max}}$ & ${\cal H}_{\mbox{\tiny max}} (\times 10^{8})$ & $L_{\mbox{\tiny min}}$ (pc) \\
\noalign{\smallskip}\hline\noalign{\smallskip}
1 & $-6588.6 \pm 31.7$ & $(-7.4,+2.2)$ & $0.045$ & $1.5$ & 32 \\
2 & $-6707.0 \pm 64.1$ & $(+5.6,+25.0)$ & N/A & N/A & N/A \\
3 & $-6610.5 \pm 43.2$ & $(-5.9,+7.2)$ & $0.043$ & $1.2$ & 41 \\
4 & $-6588.7 \pm 33.2$ & $(-7.7,+2.4)$ & $0.046$ & $1.5$ & 31 \\
Joint & $-6601.8 \pm 18.3$ & $(-3.4,+2.1)$ & $0.020$ & $0.7$ & 71 \\
\noalign{\smallskip}\hline
\end{tabular}
\end{table}
$\Delta=(r_{\mbox{\tiny NS}}-r_{\mbox{\tiny GR}})/r_{\mbox{\tiny GR}}$,
$r_{\mbox{\tiny NS}}$ is the measured north-south relativistic drift rate
(with 1$\sigma$ reported uncertainties) and
$r_{\mbox{\tiny GR}}=-6606.1$~mas/yr is the general relativity prediction
for geodetic precession given the actual GPB orbit \cite{E11}.
We use $r_0=7018$~km \cite{L07} together with recent measurements of the
Hubble parameter $H_0=74\pm 2$~km~s$^{-1}$~Mpc$^{-1}$ \cite{R11} and
normalized dark-energy density $\Omega_{\Lambda}=0.73\pm0.04$ from
high-redshift supernovae \cite{S11} and the cosmic microwave background
\cite{H12}. Our limits in each case come from the largest allowed
{\em negative\/} values of $\Delta$ in Column~3 of Table~1.

These limits are consistent with pre-GPB expectations for both the soliton
\cite{LO00} and canonical metrics \cite{OW98}.  For example, using the
joint confidence region for all four gyros and assuming that $\Lambda=3/L^2$
for the canonical metric, we find that ${\cal H}\lesssim7\times 10^7$,
implying that the component of test-body angular momentum along the
$\ell$-direction could be significantly larger than that along $r$
(modulo a $\cosh$-term). If instead we assume that ${\cal H}\sim 1$ then
$L\gtrsim 70$~pc. These limits are complementary to somewhat stronger ones
that can be obtained (with some additional assumptions) from cosmological
tests, especially the magnitude-redshift relation for distant supernovae
\cite{OWM07}. However, what is remarkable here is that there should be any
sensitivity at all to cosmological parameters such as $L$ or $\Lambda$
in an experiment involving the motions of macroscopic test bodies in
low-earth orbit (this appears to be unique to the canonical solution with 
its $(\ell/L)^2$ prefactor on the 4D part of the metric.)
Similarly, our constraints for the soliton metric are complementary to
stronger ones that can be placed on $b$ for various solar-system bodies
(including the Earth) using observational limits on violations of the
equivalence principle \cite{O00}.  The complementarity is nicely
illustrated by the case of gyro~2, whose experimental range of uncertainty
encompasses only positive values for $\Delta$ (i.e., geodetic precession
rates greater than that predicted by GR). If this result had been confirmed
by all four gyros, then the 5D theory would have been excluded using
{\em either\/} choice of metric. In the event, it is thought that gyro~2
suffered from larger systematic effects than the others \cite{W11b},
and Kaluza-Klein gravity remains a viable extension of 4D general relativity.

The question is sometimes raised whether tests like this can ever 
conclusively establish the existence or non-existence of extra dimensions.
Similar questions could perhaps be raised about other extrapolations from
the current standard model. The best approach is to apply as many
independent tests as possible and look for an accumulation of evidence
that points to the same region of parameter space in the theory.
Kaluza-Klein solitons have unique density profiles and other properties
that make them viable dark-matter candidates \cite{W94}, with theoretical
arguments suggesting values of $|b|\sim10^{-8}-10^{-2}$ in the solar system
and $|b|$ as large as $\sim0.1$ in galaxy clusters \cite{LO00}.
If two or more independent tests (such as gravitational lensing)
were to converge on a nonzero value of $b$ for the same system,
it would not {\em prove\/} the existence of extra dimensions---one might
look for scalar-field or other theories that could accommodate the same
phenomena---but it would be a strong {\em prima facie\/} case.  Moreover
the example of the canonical metric shows that it is prudent to keep an open
mind even in the case of a null result. With that metric no departure from
4D GR is expected for the classical tests; the effects of the extra
dimension manifest themselves only for spinning test bodies.
More than anything else, these remarks highlight the need for new
solutions of the field equations in $D>4$, and for a generalization
of Birkhoff's theorem that could be used to discriminate between them
in a compelling way.

Future work can build on these results in several ways.
It would be of interest to test the predictions of Kaluza-Klein gravity for
frame-dragging as well as geodetic precession. The case $k\neq0$ for the
soliton metric deserves further study. It may be possible to obtain
stronger constraints for the canonical metric using violations of the
equivalence principle \cite{O12}. Following the lead of
Matsuno and Ishihara \cite{MI09}, it might also be useful to investigate
precession effects for other 5D soliton-like solutions incorporating
time-dependence \cite{LWP93}, additional $\ell$-dependence \cite{BW96} and
electric charge \cite{LW96b,LW97}. Finally, methods similar to those
employed here can also be used to test other generalizations of 4D GR,
like those that violate Lorentz invariance \cite{O07}.

\begin{acknowledgements}
J.M.O. thanks C.W.F.~Everitt, R.J.~Adler, A.~Silbergleit and the other members
of the Gravity Probe~B theory group for discussions. R.D.E. acknowledges the Fisher College of Science and Mathematics at Towson University for travel support to present these results.
\end{acknowledgements}



\end{document}